%Paper: hep-ph/9411429
%From: jlouis@lswes8.ls-wess.physik.uni-muenchen.de
%Date: Wed, 30 Nov 94 05:24:15 +0100

%
%
%macropackage=phyzzx
%%%%%%%%10%%%%%%%%20%%%%%%%%30%%%%%%%%40%%%%%%%%50%%%%%%%%60%%%%%%%%70
%%%%%%%%80
% Format switch
%
% uncomment ONE of the following definitions
%
%\let\SELECTOR=P      % un-reduced Preprint format
\let\SELECTOR=R      % Reduced preprint format
%
% the Reduced format may not work with some DVI drivers
%%%%%%%%%%%%%%%%%%%%%%%%%%%%%%%%%%%%%%%%%%%%%%%%%%%%%%%%%%%%%%%%%%%%%%
%%%
%
% Macros for the paper
%%%%%%%%%%%%%%%%%%%%%%%%%%%%%%%%%%%%%%%%%%%%%%%%%%%%%%%%%%%%%%%%%%%%%%
%%%
% Load PHYZZX
%
\expandafter\ifx\csname phyzzx\endcsname\relax \input phyzzx \fi
%
% general setup
%
\interdisplaylinepenalty=10000
%\VOFFSET=-0.5in
%
% Set oneup and two-up formats
%
\newdimen\doublewidth
\doublewidth=12in
\newinsert\LeftPage
\count\LeftPage=0
\dimen\LeftPage=\maxdimen
\def\PageBox{\vbox{\makeheadline \pagebody \makefootline }}
\def\papersize{\hsize=412pt \vsize=570pt \pagebottomfiller=0pt
    \skip\footins=\bigskipamount \normalspace }
\papersize
\if R\SELECTOR
    \mag=833
    \voffset=-0.3truept
    \hoffset=-0.5truein
    \output={\ifvoid\LeftPage \insert\LeftPage{\floatingpenalty 20000
\PageBox}
        \else \shipout\hbox to\doublewidth{%
            \box\LeftPage \hfil \PageBox }\fi
        \advancepageno
        \ifnum\outputpenalty>-20000 \else \dosupereject \fi }
    \message{Warning: some DVI drivers cannot handle reduced
output!!!}
\else
    \mag=1000
    \voffset=\VOFFSET
    \hoffset=0pt
\fi
%%%%%%%%%%%%%%%
\overfullrule=0pt
\def\ra{\rightarrow}
\def\O{{\cal O}}
\def\del{\partial}
\def\Mgut{M_{\rm GUT}}

\def\MX{M_X}
\def\Mstring{M_{\rm String}}

\def\alphaX{\alpha_{X}}
\def\Mweak{M_{Z}}

\def\K{K\"ahler}
\def\Ktot{{K}}
\def\Kmod{\hat K}

\def\matter{Q}
\def\matterb{\smash{\overline\matter}\vphantom{\matter}}
\def\mod{M}
\def\modb{\smash{\overline M}\vphantom{M}}
\def\Fb{\smash{\overline F}\vphantom{F}}

\def\jb{{\bar\jmath}}
\def\Ib{{\bar I}}
\def\Jb{{\bar J}}
\def\ind{\phi}
\def\indb{\bar{\phi}}
\def\mzero{m^{2}_0}
\def\Azero{A_0}

\def\hh{h}
 
\def\Re{\mathop{\rm Re}\nolimits}

\interdisplaylinepenalty=10000

%W's
%

\def\Wnp{\hat W}

%
%masses
%
\def\squark{\tilde{q}}

\def\down{\tilde{d}}
\def\slepton{\tilde{\ell}}
\def\mg{\tilde m}
\def\mgzero{\tilde{m}_{1/2}}
\def\mgrav{m_{3/2}}
\def\msquark{m_{\squark}}
\def\mslepton{m_{\slepton}}
\def\mav{m}
\def\MM{M}
\def\msf{m_{\tilde f}}

\REF\MSSM{For a review see for example, %\brk
H.-P.~Nilles, Phys.~Rep.~C110 (1984) 1;\brk
H.E.~Haber and G.~Kane,  Phys.~Rep.~C117 (1985) 75;\brk
R.~Barbieri, Riv.~Nuovo Cimento 11 (1988) 1;\brk
L.E.~Ib\'a\~nez and G.G.~Ross,
 in {\it Perspectives in Higgs Physics}, ed.~ G.~Kane;\brk
F.~Zwirner, preprint CERN-TH.6357/91, Talk at the Workshop on Physics
and Experiments with Linear Colliders, Saariselka, Finland,
Sep.~1991;\brk
and references therein.}
\REF\BG{P.~Bin\'etruy and M.K.~Gaillard, Nucl.~Phys.~B358 (1991) 121.}
\REF\CFILQ{M.~Cveti\v c, A.~Font, L.E.~Ib\'a\~nez, D.~L\"ust and
F.~Quevedo, Nucl.~Phys.~B361 (1991) 194.}
\REF\IL{L.~Ib\'a\~nez and D.~L\"ust, Nucl.~Phys.~B382 (1992) 305.}
\REF\CCM{B.~de Carlos, J.A.~Casas and C.~Mu\~noz,
Phys.~Lett.~B299 (1993) 234.}
\REF\MR{A.~de la Macorra and G.G.~Ross, Nucl.~Phys.~B404 (1993) 321.}
\REF\KL{V.S.~Kaplunovsky and J.~Louis, Phys.~Lett.~B306 (1993) 269.}
\REF\BIM{A.~Brignole, L.E.~Ib\'a\~nez and C.~Mu\~noz,
 Nucl.~Phys.~B422 (1994) 125.}
\REF\FCNC{For early works see for example,\brk
S.~Dimopoulos and H.~Georgi, Nucl.~Phys.~B193 (1981) 150;\brk
J.~Ellis and D.V.~Nanopoulos, Phys.~Lett.~110B (1982) 44;\brk
R.~Barbieri and R.~Gatto, Phys.~Lett.~110B (1982) 211;\brk
M.~Duncan, Nucl.~Phys.~B221 (1983) 285;\brk
J.~Donoghue, H.-P.~Nilles, and D.~Wyler, Phys.~Lett.~B128 (1983)
55;\brk
A.Bouquet, J.~Kaplan and C.A.~Savoy, Phys.~Lett.~B148 (1984) 69.}
\REF\HKR{L.J.~Hall, V.A.~Kostelecky and S.~Raby,
 Nucl.~Phys.~B267 (1986) 415;\brk
H.~Georgi, Phys.~Lett.~B169 (1986) 231.}
\REF\HLW{L.J.~Hall, J.~Lykken and S.~Weinberg,
 Phys.~Rev.~D27 (1983) 2359.}
\REF\DKS{M.~Dine, A.~Kagan and S.~Samuel, Phys.~Lett.~B243 (1990) 250.}
\REF\DiNe{
M.~Dine and W.~Fischler, Phys.~Lett.~110B (1982) 227;\brk
M.~Dine and A.E.~Nelson, Phys.~Rev.~D48 (1993) 1277.}
\REF\DLK{M.~Dine, A.~Kagan and R.~Leigh, Phys.~Rev.~D48 (1993) 4269.}
\REF\NS{Y.~Nir and N.~Seiberg, Phys.~Lett.~B309 (1993) 337.}
\REF\LNS{M.~Leurer, Y.~Nir and N.~Seiberg, Nucl.~Phys.~B420 (1994)
468.}
\REF\MN{A.~Lleyda and C.~Mu\~noz, Phys.~Lett.~B317 (1993) 82;\brk
N.~Polonsky and A.~Pomarol, Phys.~Rev.~Lett.~73 (1994) 2292;\brk
T.~Kobayashi, D.~Suematsu and Y.~Yamagishi,  Phys.~Lett.~B329 (1994)
 27;\brk
D.~Matalliotakis and H.P.~Nilles, Munich preprint TUM-HEP-201/94;\brk
M.~Olechowski and S.~Pokorski, Munich preprint MPI-PHT/94-40;\brk
T.~Kobayashi, D.~Suematsu, K.~Yamada and Y.~Yamagishi,
preprint KANAZAWA-94-16;\brk
P.~Brax and M.~Chemtob, Saclay preprint SACLAY-SPHT-94-128.}
\REF\CEKLP{D.~Choudhury, F.~Eberlein, A.~K\"onig, J.~Louis and
 S.~Pokorski, Munich preprint LMU--TPW 94-12.}
\REF\noscale{E.~Cremmer, S.~Ferrara, C.~Kounnas and D.~Nanopoulos,
Phys.~Lett.~133B (1983) 61; \brk
for a review see, A.B.~Lahanas and D.V.~Nanopoulos,
Phys.~Rep.~C145 (1987) 1. }
\REF\EDMN{For early works see for example,\brk
J.~Ellis, S.~Ferrara and D.V.~Nanopoulos,
 Phys.~Lett.~114B (1982) 231;\brk
W.~Buchm\"uler and D.~Wyler, Phys.~Lett.~121B (1983) 321;\brk
J.~Polchinski and M.B.~Wise, Phys.~Lett.~125B (1983) 393.}
\REF\AGNT{I.~Antoniadis, E.~Gava and K.~Narain,
 Phys.~Lett.~B283 (1992) 209, Nucl.~Phys.~B383 (1992) 93;\brk
I.~Antoniadis, E.~Gava, K.~Narain and T.~Taylor,
Nucl.~Phys.~B407 (1993) 706.}
\REF\KN{J.K.~Kim and H.-P.~Nilles, Phys.~Rev.~Lett.~73 (1994) 1758.}
\REF\FKZ{S.~Ferrara, C.~Kounnas and F.~Zwirner,
CERN preprint CERN-TH-7192-94.}
\REF\DKLb{L.~Dixon, V.~Kaplunovsky and J.~Louis,
 Nucl.~Phys.~B355 (1991) 649.}
\REF\BLM{R.~Barbieri, J.~Louis and M.~Moretti,
 Phys.~Lett.~B312 (1993) 451, \brk erratum ibid.~B316 (1993) 632.}
\REF\LNZ{J.~Lopez, D.~Nanopoulos and A.~Zichichi, Phys.~Lett.~B319
(1993) 451.}
\REF\PDG{Particle Data Group, Phys.~Rev.~D50 (1994) 1173.}
\REF\GM{F.~Gabbiani and A.~Masiero, Nucl.~Phys.~B322 (1989) 235.}
\REF\HKT{J.~Hagelin, S.~Kelley, T.~Tanaka, Nucl.~Phys.~B415 (1994)
293;\brk Mod.~Phys.~Lett.~A8 (1993) 2737;\brk
T.~Kosmas, G.K.~Leontaris and J.D.~Vergados,
 Phys.~Lett.~219B (1989) 457,
 Prog. Part. Nucl. Phys.~33 (1994) 397; \brk
G.K.~Leontaris, Z. Phys.~C62 (1994) 91.}
\REF\DGH{M.~Dugan, B.~Grinstein and L.~Hall,
    Nucl.~Phys.~B255 (1985) 413.}
\REF\FPT{W.~Fischler, S.~Paban and S.~Thomas,
 Phys.~Lett.~B289 (1992) 373.}
\REF\BV{S.~Bertolini and F.~Vissiani, Phys.~Lett.~B324 (1994) 164.}
\REF\AlSm{I.~Altarev {\it et al.}, JETP Lett.~44 (1986) 460;
 K.~Smith {\it et al.}, Phys.~Lett.~B234 (1990) 191.}
\REF\Abdu{K.~Abdullah {\it et al.}, Phys.~Rev.~Lett.~65 (1990) 2347.}
\REF\Jaco{J.P.~Jacobs {\it et al.}, Phys.~Rev.~Lett.~71 (1993) 3782.}
\REF\Japb{T.~Kobayashi, M.~Konmura, D.~Suematsu, K.~Yamada and
Y.~Yamagishi, preprint KANAZAWA-94-17.}
\REF\choi{K.~Choi, Phys.~Rev.~Lett.~72 (1994) 1592.}
\REF\BaGi{R.~Barbieri and G.F.~Giudice, Nucl.~Phys.~B306 (1988) 63.}
\REF\DKL{L.J.\ Dixon, V.S.\ Kaplunovsky and J.\ Louis,
  Nucl.~Phys.~B329 (1990) 27.}
\REF\CFKRG{G.D.~Coughlan, W.~Fischler, E.W.~Kolb, S.~Raby and
 G.G.~Ross, \brk Phys.~Lett.~131B (1983) 59.}
\REF\BKN{T.~Banks, D.~Kaplan and A.~Nelson, Phys.~Rev.~D49 (1994)
779.}
\REF\CCQR{B.~de Carlos, A.~Casas, F.~Quevedo and E.~Roulet,
 Phys.~Lett.~B318 (1993) 447.}
\REF\BCMN{T.~Banks, A.~Cohen, G.~Moore and Y.~Nir, unpublished.}
\REF\RaTo{L.~Randall and S.~Thomas, preprint MIT-CTP-2331.}
%%%%%%%%%%%%%%
\def\Weizmann{\centerline{\it Department of Particle Physics}
  \centerline{\it Weizmann Institute of Science, Rehovot 76100,
Israel}}
\def\Munich{\centerline{\it Sektion Physik, Universit\"at M\"unchen}
\centerline{\it Theresienstr.~37, D-80333 M\"unchen, Germany}}
{\baselineskip=11pt
\Pubnum={LMU-TPW 94--17 \cr WIS-94/50/Dec-PH}
\date={December, 1994}
\titlepage
\title{{\bf Some Phenomenological Implications of String Loop Effects}}
\author{Jan Louis}
\Munich
\andauthor{Yosef Nir}
\Weizmann
\vskip .5cm
\centerline{\bf Abstract}
We investigate  low energy implications of
string loop corrections to supergravity couplings
which  break a possible flavor universality of the tree level.
If Supersymmetry is broken by the dilaton $F$-term, universal soft
scalar masses arise at the leading order but string loop corrections
generically induce flavor-non-diagonal soft terms. Constraints from
flavor changing neutral currents (FCNC) and CP violation then require
a large supersymmtery breaking scale and thus heavy gluinos and squarks.
If Supersymmetry is broken by moduli $F$-terms,
universality at the string tree level can only be guaranteed by
extra conditions on the K\"ahler potential.
A large hierarchy between the gluino and squark masses ensures that
FCNC and CP violation constraints are satisfied.
If the soft scalar masses vanish at the string tree level,
the cosmological problems related to light moduli can be evaded.
However, generic string  loop corrections  violate
FCNC bounds and require very heavy squark masses ($\approx 100\, TeV$).
\endpage
}

%%%%%%%%%%%%%%%%%%%%%%%%%%
\chapter{Introduction}
Supersymmetric extensions of the Standard Model (SM) are among
the most promising candidates for new physics  above
the weak scale $\Mweak$.
In the minimal supersymmetric SM (MSSM) all particles of the
SM are promoted to chiral $N=1$ supermultiplets
with one additional Higgs doublet added [\MSSM].
Supersymmetry is assumed to be broken by explicit but soft breaking
terms which appear  naturally in the low energy limit
of spontaneously broken supergravity theories.
This soft supersymmetry breaking introduces a number of new
parameters into the Lagrangian which control
the mass spectrum of the fermion's superpartners.
The parameter space spanned by the soft terms
has been the subject of numerous investigations,
mostly under some simplifying assumptions [\MSSM].
It has also been  studied
within the context of  further extensions such as superstring
theory [\BG --\BIM].

Without specifying the precise origin of the non-perturbative
effects in superstring theory but with
 a number of plausible assumptions, it has been possible
to observe interesting features of the soft terms [\IL, \KL, \BIM].
In particular, many signatures at $\Mweak$ are entirely controlled
by perturbative couplings in string theory and (almost) independent
of the assumption about the unknown non-perturbative properties.
The relevant perturbative couplings have indeed been computed
for many string vacua at the leading order (tree level)
in string perturbation theory.
One finds that in most cases the standard assumptions
of the MSSM are not  fulfilled. For example,
non-universal scalar masses as well as non-proportional
$A$-terms with arbitrary CP-violating phases are easily generated.
Non-universal scalar masses are particularly dangerous
since they generically induce unacceptably large contributions
to rare processes such as flavor changing neutral currents
(FCNC) [\FCNC--\DKS]. One way out is to look for possible
mechanisms which naturally suppress non-universal scalar masses.
This could be natural if SUSY breaking is communicated to the light
particles by gauge interactions [\DiNe] or in models with a
nonabelian
horizontal symmetry [\DLK]. (Abelian horizontal symmetries could
align quark and squark mass matrices, thus suppressing FCNC without
squark degeneracy [\NS,\LNS].)\foot{Other recent investigations
of the problem include refs.~[\MN, \CEKLP].}

In the context of string theory, universality
of squark masses is achieved if the dominant source for supersymmetry
breaking is the dilaton $F$-term [\KL] or in `no-scale' type of
models [\noscale].
However, in both cases non-universal scalar masses might arise
through couplings generated at the 1-loop level of string perturbation
theory. Although for such couplings much less (string) information is
currently available, it is possible to estimate the typical size of
these corrections and hence estimate the physical implications for weak
scale phenomenology. Furthermore, the generic CP-violating phases
of the $A$ and $B$-terms are constrained in both scenarios
and can be confronted with the bounds for
the electric dipole moment of the neutron (EDMN)[\EDMN].

This paper is organized as follows: In Section 2 we present the
general form of string loop corrections. Their implications on FCNC and
CP violation when supersymmetry is broken by the dilaton $F$-term
are studied in section~3.
A similar analysis for supersymmetry breaking by  moduli $F$-terms
(including some cosmological consequences) are studied  in section~4.
A summary is given in Section~5.

%%%%%%%%%%%%%%%%%%%%%%%%%%%%%%%%%%%%%
\chapter{String Loop Corrections}
We first summarize the generic structure of
the couplings in the low energy effective Lagrangian of string
theory. In addition to the gravitational and gauge multiplets,
the massless spectrum contains two types of  chiral supermultiplets.
First, matter fields $\matter^I$ which are charged under the
low energy gauge group $G$ and which contain the quark and lepton
multiplets of the SM. Second, there are the gauge neutral
supermultiplets $S$ (dilaton) and $\mod^i$ (moduli)
which are flat directions of the perturbative effective potential and
whose VEVs parameterize the perturbative degeneracy of the
string vacuum.\foot{Strictly speaking there can also be
singlet supermultiplets which are not moduli,
 \ie\ which are not a flat direction of the effective
potential. For the purpose of this article we include
them among the matter fields $\matter^I$.}  The couplings of the
low energy effective Lagrangian for the massless multiplets
 are encoded in three scalar functions:
the real \K\ potential $\Ktot$, the holomorphic superpotential
$W$ and the holomorphic gauge kinetic function $f$.

$K$ summarizes the kinetic energy terms and
at low energies can be expanded in the matter fields
$$
\Ktot\ =\ \kappa^{-2}\, \Kmod\ +\ Z_{\Ib J}
\matterb^\Ib e^{2V} \matter^J\ +\ ({1\over2}H_{IJ}
Q^IQ^J+{\rm h.c.})\ +\ \cdots \ ,
\eqn\Kexpansion$$
where the `$\cdots$' in eq.~\Kexpansion\ correspond to terms which
are  irrelevant for the present investigation.
The matter fields $\matter^I$ carry canonical dimension one whereas
$S$ and $\mod^i$ are expected to receive Planck-sized VEVs and
therefore are chosen to be dimensionless. The couplings $\Kmod$,
$Z_{\Ib J}$ and $H_{IJ}$ are dimensionless functions of  $S$
and $\mod^i$ and only further constrained by the fact that the
dilaton $\Re S$ serves as the string-loop counting parameter. At the
string tree level the dilaton couples universally in all
string vacua;  this universality is lost at the loop level but
all couplings can be expanded in powers of  $\Re S$:
$$
\eqalign{
\Kmod \  &=\ -  \ln (S + \overline S)\ +\
\sum_{n=0}^\infty  {\Kmod^{(n)} (\mod,\modb)\ \over\,
[ 8\pi^2 (S + \overline S )]^n}  \, , \cr
Z_{\Ib J}\ &=\ \sum_{n=0}^\infty  {Z^{(n)}_{\Ib J} (\mod,\modb)\
\over\, [ 8\pi^2 (S + \overline S )]^n} \, ,  \cr
H_{I J}\ &=\ \sum_{n=0}^\infty  {H^{(n)}_{I J} (\mod,\modb)\ \over\,
[8\pi^2 (S + \overline S )]^n}  \, , }
\eqn\dilatonkol
$$
where
$\Kmod^{(n)}$, $Z^{(n)}_{\Ib J}$ and $H^{(n)}_{IJ}$ do not depend
on the  dilaton and  their moduli dependence  in general cannot be
further constrained
(they do depend  on the details of the internal superconformal
field theory).\foot{$\Kmod^{(1)}$ and $Z^{(1)}_{\Ib J}$
are the four-dimensional analogue of the Green-Schwarz term and have
recently been computed in some orbifold vacua [\AGNT].}

The scalar potential and the Yukawa
couplings $\tilde Y_{IJL}$ are determined
by the superpotential $W$ which is not renormalized
at any order in string perturbation theory.
The perturbative $W$ is completely independent of the dilaton $S$
but non-perturbative  corrections can introduce
further dilaton (and moduli) dependence into $W$.
Expanding  in $\matter^I$ we have
$$
W=\Wnp(S, \mod^i)+{1\over2}\tilde\mu_{IJ}(S, \mod^i)Q^IQ^J+
{1\over3}\tilde Y_{IJL}(\mod^i)Q^IQ^JQ^L+\cdots.
\eqn\Super
$$
where the `$\cdots$' stand for non-renormalizable interactions.
$\Wnp$ is identically  zero at any order in string perturbation
theory and arises only from  non-perturbative physics.
(Similarly, the $S$-dependence in $\tilde\mu$ is induced at the
non-perturbative level.)
Without specifying the precise nature of such non-perturbative
effects they can be  parameterized by $\Wnp$.
We assume that $\Wnp$  is such that it breaks
supersymmetry by generating non-vanishing
moduli $F$-terms $\vev{F^i}$ and/or a dilaton $F$-term $\vev{F^S}$.
To simplify our notation let us introduce an index $\ind$
which runs over both the moduli and dilaton direction, \ie\
$\ind = (i, S)$. Using this notation the $F$-terms are given by
$$
\Fb^{\bar \ind} = \kappa^2 e^{\Kmod/2} \Kmod^{\indb \ind}
(\del_{\ind} \Wnp + \Wnp \del_\ind \Kmod)\ ,
\eqn\fterms$$
while the scale of supersymmetry breaking is  parameterized
by the (complex) gravitino mass
$$m_{3/2}=\kappa^2 e^{ \hat K/2}\hat W.\eqn\mth$$
We further assume that, at the minimum of the potential,
a dilaton VEV $\vev S$ and moduli VEVs $\vev{\mod^i}$
are generated and hence the perturbative vacuum degeneracy is
(partially) lifted. Finally, the cosmological constant is assumed to
be zero which implies
\foot{
Recently various mechanisms have been studied which also include
low energy quantum corrections to the cosmological
constant [\KN,\FKZ]. Most of our  analysis here is
insensitive to the details of the mechanism responsible for the
vanishing of the cosmological constant.}
$$
|m_{3/2}|^2 =\coeff13  \hat{K}_{\ind \indb} F^\ind \Fb^{\indb}  .
\eqn\cosmo$$

Under these assumptions (spelled out in more detail in ref.~[\KL]) soft
supersymmetry breaking terms are generated in the observable sector.
In particular, the potential for the observable matter scalars (which
we also call $Q^I$) contains the following soft supersymmetry  breaking
terms:
$$
V^{({\rm SSB})}=m_{I\Jb}^2 Q^I \matterb^{\Jb}\ +\
({1\over3}A_{IJL}Q^IQ^JQ^L+{1\over2}B_{IJ}Q^IQ^J+{\rm h.c.}),
\eqn\VSSB$$
where the parameters $m^2, A, B$ are moduli and dilaton dependent and
not necessarily flavour diagonal [\IL, \KL, \BIM]. Specifically,
$$m^2_{I\bar J}=|m_{3/2}|^2Z_{I\bar J}-F^\ind\overline F^{\indb  }
R_{\ind\indb   I\bar J},\eqn\MsIJ$$
where the flavour dependence can arise through the (perturbative)
curvature couplings
$$R_{\ind \indb I\bar J}=\partial_\ind \bar\partial_{\indb}Z_{I\bar
J}-
\Gamma^N_{\ind I}Z_{N\bar L}\bar\Gamma^{\bar L}_{\indb  \bar J},\ \ \
\Gamma^N_{\ind I}=Z^{N\bar J}\partial_\ind  Z_{\bar JI},\eqn\RGamma$$
and hence the standard assumption of universal (flavour independent)
soft masses might not hold.  Furthermore,
$$
A_{IJL}=F^\ind (\partial_\ind Y_{IJL}-\Gamma^N_{\ind (I}Y_{JL)N}
+{1\over2}\hat K_\ind Y_{IJL}) \ ,
\qquad Y_{IJK} = e^{\hat K/2}\tilde Y_{IJK}\ ,
\eqn\AIJL
$$
where the first two terms are in general not proportional to the
Yukawa couplings. Similarly,
$$\eqalign{
B_{IJ}=&2|m_{3/2}|^2H_{IJ}+m_{3/2}F^\ind D_\ind H_{IJ}-\bar m_{3/2}
\overline F^{\bar \ind }\bar\partial_{\bar \ind }H_{IJ}-
\overline F^{\bar \ind}
F^\ind D_\ind\bar\partial_{\bar \ind }H_{IJ}\cr +&e^{\hat K/2}
[F^\ind (\partial_\ind \tilde\mu_{IJ}+\hat K_\ind \tilde\mu_{IJ}
-2\Gamma^K_{I\ind }\tilde\mu_{KJ})-\bar m_{3/2}\tilde\mu_{IJ}],\cr
}\eqn\muBIJ$$
(where $D_\ind H_{IJ}=\partial_\ind H_{IJ}-2\Gamma^K_{I\ind }H_{KJ}$)
is not necessarily proportional to
\foot{In the context of the MSSM there is only one $B$-term
allowed by gauge invariance and $R$-invariance and hence no flavour
dependent matrix exists. What the non-proportionality means in this case
is that $\mu$ can be zero with $B$ staying finite (or vice versa).}
$$
\mu_{IJ}=e^{\hat K/2}\tilde\mu_{IJ}+m_{3/2}H_{IJ}-
\overline F^{\indb  }\bar\partial_{\indb  }H_{IJ}\ .
\eqn\muIJ
$$
Hence, the parameters of  $V^{({\rm SSB})}$ in eq.~\VSSB\ in general
do not  satisfy the property of flavour-independence
which is commonly assumed in the MSSM.

Finally, there is one more soft term induced:
the gauginos aquire a mass given by
$$
\mg_a = F^\ind \del_\ind \ln g_a^{-2}\ ,
\eqn\gauginomass$$
where $g_a^{-2}$ are the gauge couplings ($a$ labels the simple
factors in the gauge group).
In string theory the gauge couplings  are universal
at the leading order and determined by the VEV of the dilaton.
Non-universality and moduli dependence is only introduced
via  one-loop threshold corrections $\Delta_a$ [\DKLb]
$$
g_a^{-2}(\Mstring)= \Re S + {\Delta_a (\mod,\modb)\over 16\pi^2}\ .
\eqn\gaugecoup$$
$\Mstring$ denotes the characteristic scale of string theory;
numerically $\Mstring \approx 5\times 10^{17}$ GeV which is close to
the supersymmetric GUT-scale $\Mgut\approx 3\times 10^{16}$ GeV.
In this paper we do not make any distinction between the two scales
and denote them both by $\MX$.
As a consequence of the very special field dependence of
the gauge couplings, the gaugino masses  are universal
at the leading order and obey
$$
\mg_a = \mgzero + {\alphaX\over 4\pi} \mg^{(1)}_a + \cdots\ ,
\eqn\gmassim $$
where
$$
\eqalign{
\mgzero =\ & {F^S\over (S + \overline S)}\ , \qquad
 \mg^{(1)}_a = F^i \del_i \Delta_a - F^S \Delta_a\, , \cr
\alphaX =\ &  {g^2 (\MX) \over 4\pi} = {1\over 2\pi (S+\overline S)}\ .}
\eqn\mgapprox $$
Note that the universal gaugino mass $\mgzero$ is directly proportional
to the dilaton $F$-term $F^S$ and that both $\mgzero$ and
$\mg^{(1)}_a$ are of order $\O(\mgrav)$.

On the other hand, the scalar masses given by eq.~\MsIJ\
are in general flavour-dependent (non-universal)
already at the leading order of perturbation theory
when $Z_{I\bar J}$ is approximated by its tree level
contribution $Z_{I\bar J}^{(0)}$.
However, there are scenarios where universal scalar masses
and $A$-terms do appear at the leading order
and non-universality is only introduced at the one-loop level.
For those cases - which are the focus of this paper - we have
$$
\eqalign{
m^2_{I\Jb} =&\ \mzero\,  Z_{I\bar J}^{(0)}
              + {\alphaX\over 4\pi}\, m^{2\, (1)}_{I\Jb}
              + \cdots\ , \cr
A_{IJL} =&\ \Azero\, Y_{IJL}
              + {\alphaX\over 4\pi}\, A^{(1)}_{IJL}+ \cdots\ . \cr}
\eqn\softexp $$
%
%
%%%%%%%%%%%%%%%%%%%%%%%%%%%%%%%%%%%%%%%%%%%%%%%%%%%%%%%%
\chapter{Supersymmetry Breaking by the Dilaton}
Under the assumption that only a dilaton $F$-term $\vev{F^S}$
is generated by the non-perturbative physics, the soft parameters
simplify considerably at the leading order and the scalar masses
and $A$-terms are indeed universal. This is a consequence
of the universal couplings of the dilaton at the string tree level.
Specifically one finds  [\KL, \BIM]
$$
\mzero =  |m_{3/2}|^2 = \coeff13 {|F^S|^2\over  (S + \overline S)} ,\qquad
\Azero = - {F^S \over (S + \overline S)}, \qquad
\mgzero = {F^S \over (S + \overline S)}, \eqn\FS
$$
while $B$ and $\mu$ are independent parameters.
(If, in addition, $\tilde \mu =0$ holds in eq.~\muIJ, $B$ and $\mu$
are related via $B = 2\, \bar{m}_{3/2}\, \mu$
but we do not assume this relation here.)
Given the soft terms \FS\ generated at $\MX$, standard
RG-analysis can be used to compute the supersymmetric mass spectrum
at low energies [\BLM].\foot{See also ref.~[\LNZ].}
One finds that all squark masses $\msquark$ are
essentially degenerate with the gluino mass $\mg_3$
$$
\msquark \simeq \mg_3 \simeq 5\, \mgrav\ ,
\eqn\squarks $$
whereas the slepton masses obey
$$
\mslepton \simeq 0.3\, \mg_3 \simeq 1.5\,  \mgrav\ .
\eqn\sleptons $$
In order to evade the direct experimental bounds [\PDG]
on scalar and gaugino masses, eqs.~\squarks, \sleptons\ imply
$$
\mgrav > 30\, {\rm GeV}\ .
\eqn\mbound $$

In this section we study the physical properties of this scenario
beyond the leading order approximation. In particular, we assume
 generic $\O(1)$ one-loop couplings $Z^{(1)}_{I\Jb}$
which induce flavour-dependent
scalar masses  and non-proportional $A$-terms  at the next order.
{}From eqs.~\MsIJ-\AIJL\ we learn (using \FS)
$$
\eqalign{
m^{2\, (1)}_{I\Jb} =&-5\ |m_{3/2}|^2 Z^{(1)}_{I\Jb}\sim
\O(\mgrav^2),\cr
A_{IJL}^{(1)} =&{F^S\over(S+\overline S)}\left(-\hat{K}^{(1)}\, Y_{IJL}
+ 3 Z^{(1)}_{I\Jb} Z ^{\Jb N \, (0)}\, Y_{NJL}
+ \del_\jb \hat{K}^{(1)} \hat{K}^{(0)\, \jb i} D_i Y_{IJL}\right)\cr
\sim & \O(\mgrav Y), \cr
}
\eqn\FScorr $$
where $Z^{(0)},Z^{(1)}$ and $\hat K^{(1)}$ are all functions of $\O(1)$.

\section{Constraints from Flavor Changing Neutral Currents}
Let us first focus on the constraints implied by the smallness of
FCNC. We use the notation of ref.~[\NS] and the calculations of
ref.~[\GM].\foot{See also refs.~[\HKT].}
The experimental bounds from FCNC constrain the
sfermion masses at the weak scale $\MM^{f2}$ ($f=u,d,\ell$)
which are determined in terms of the soft
input parameters \FS\ and \FScorr\ generated at the high energy
scale $\MX$. In the basis where fermion mass-matrices are
diagonal and gaugino couplings are diagonal, the sfermion masses
appear in $3\times3$ submatrices,
$$\MM^{f2}=\pmatrix{\MM^{f2}_{LL}&\MM^{f2}_{LR}\cr
\MM^{f2}_{RL}&\MM^{f2}_{RR}\cr},
\eqn\sfmasses$$
where the soft scalar masses $m^2_{I\Jb}$  contribute to the
diagonal blocks $M^{f2}_{LL}, M^{f2}_{RR}$   while
the $A$-terms directly determine $M^{f2}_{LR}$ [\MSSM].
\foot{The diagonal elements in $\MM_{LR}^2$ also depend on $\mu$.}
As SUSY breaking by the dilaton leads to approximately
degenerate sfermions in each sector, it is convenient
to define the average sfermion mass-squared, $\msf^2$.
FCNCs are then proportional to
$$(\delta^f_{MN})_{ij}={(\MM^{f2}_{MN})_{ij}\over \msf^2}\ ,
\quad i\neq j\ \
\eqn\deltaf$$
and the strongest constraints on non-universality arise
from the light generations.
(For squarks the bounds are particularly strong on  the combination
$\VEV{\delta^f_{12}}=\sqrt{(\delta^f_{LL})_{12}(\delta^f_{RR})_{12}}$.)
One finds [\GM, \NS]
$$\eqalign{{\rm Re}\VEV{\delta^d_{12}}\leq6\times10^{-3}
\left({\mav_{\down}\over1\ TeV}\right),&\quad
{\rm Re}(\delta^d_{LR})_{12}\leq8\times10^{-3}
\left({\mav_{\down}\over1\ TeV}\right)\ ;\cr
{\rm Im}\VEV{\delta^d_{12}}\leq5\times10^{-4}
\left({\mav_{\down}\over1\ TeV}\right),&\quad
{\rm Im}(\delta^d_{LR})_{12}\leq7\times10^{-4}
\left({\mav_{\down}\over1\ TeV}\right),\cr
}\eqn\KBDmixMM$$
from $\Delta m_K$ and $\epsilon_K$, and
$$
(\delta^\ell_{MM})_{12}\leq1.5\times10^{-2}\left({\mav_{\slepton}
\over0.3\ TeV}\right)^2,\quad (\delta^\ell_{LR})_{12}\leq5\times
10^{-6}\left({\mav_{\slepton}\over0.3\ TeV}\right),
\eqn\megMM$$
from the bound on BR$(\mu\rightarrow e\gamma)$. The bounds
\KBDmixMM\ and \megMM\ have been evaluated under the assumption
$\msquark\simeq\mg_{3}$ and $m_{\tilde\ell} \simeq 2\, \mg_{1}$
as appropriate for dilaton-induced SUSY breaking (\cf\ eq.~\squarks).
In the slepton sector the bound on $(\delta^\ell_{MM})_{12}$
also depends on
$(\MM^{\ell2}_{LR})_{22}\approx m_\mu\left[{A_{\mu\mu H_d}\over
Y_{\mu\mu H_d}}+\mu^*{\VEV{H_u}^*\over\VEV{H_d}}\right]$
and we have used the value
$(\MM^{\ell2}_{LR})_{22} = -3.8\, m_\mu m_{3/2}$ in \megMM\
as a characteristic value for the dilaton scenario.
\foot{$(\MM^{\ell2}_{LR})_{22}$ depends on a phase
$\phi_B$, defined in the next section. Here we take
$\phi_B=0$ which gives the weakest constraint.}
All bounds quoted are only accurate up to factors of $\O(1)$
due to hadronic uncertainties in the squark sector and the
dependence on $(\MM^{\ell2}_{LR})_{22} $ in the slepton sector.
Finally, the bounds from $\Delta m_B$, $\Delta m_D$ and radiative
$\tau$ decays are much milder than \KBDmixMM\ and \megMM\
and  play no role in our analysis.

The experimental bounds \KBDmixMM\ and \megMM\ can now be
compared with the theoretical `predictions' of the dilaton scenario
which follow from eqs.~\FS\ and \FScorr.
Let us first note that even for the universal boundary
conditions \FS\ renormalization effects induce small $\delta$'s
at low energies which obey  \KBDmixMM\ and \megMM\ [\GM, \HKT].
The point we want to study here is the implication of the
non-universality implied by \FScorr. The running of the off-diagonal
mass-matrix elements of the first two generations
is neglibly small [\MSSM] and hence we can estimate at the weak scale:
$$\eqalign{
(\delta^q_{MM})_{12}&\simeq {\alphaX\over4\pi}
{m^{2\, (1)}_{12}\over \msquark^2}\simeq 1.2\times10^{-4},\cr
(\delta^d_{LR})_{12}&\simeq \ 3\, {\alphaX\over4\pi }
{m_s \mgrav \over \msquark^2}
\simeq4\times10^{-7}\left({1\ TeV\over\msquark}\right),\cr
(\delta^\ell_{MM})_{12}&\simeq{\alphaX\over4\pi}
{m^{2\, (1)}_{12}\over \mslepton^2}\simeq 1.5\times10^{-3},\cr
(\delta^\ell_{LR})_{12}&\simeq 1.5\, {\alphaX\over4\pi }
{m_\mu\mgrav\over \mslepton^2}
\simeq1\times10^{-6}\left({0.3\ TeV\over\mslepton}\right),\cr }
\eqn\deltas$$
where we used eqs.~\squarks, \sleptons,  \FScorr\ and $\alphaX=1/24$.
Also \deltas\ are only order of magnitude estimates and  factors
of $\O(1)$ are neglected.

Comparing \deltas\ with \KBDmixMM, \megMM\ we find that
in the squark sector the only potentially interesting bound arises
from $\epsilon_K$. Assuming phases of $\O(1)$ in the mass matrix
\sfmasses\ or equivalently ${\rm Im}(\delta^d_{MM})_{12}\approx
{\rm Re}(\delta^d_{MM})_{12}$, we find that \KBDmixMM\
can  be satisfied by slightly raising $\mav_{\down}$,
$$\mav_{\down}\geq180\ GeV\ \Longrightarrow\ \mg_3\geq180\ GeV
\quad (\mgrav\geq 36\ GeV).
\eqn\mgq$$
In the slepton sector
the constraint is  stronger and \megMM\ can only be satisfied for
$$\mav_{\slepton}\geq 135\ GeV\ \Longrightarrow\ \mg_3\geq450\ GeV
\quad(\mgrav\geq 90\ GeV).
\eqn\mgl$$
The fact that the stronger constraint arises in the slepton sector is
a consequence of the large renormalization effect in the squark sector
due to the gluino mass which enhances the average squark masses and
therefore weakens the FCNC constraints [\DLK, \BIM, \CEKLP].

To summarize, when SUSY is broken by the dilaton, universality
and proportionality are violated
at the string loop level. The effect on mass differences
in the various neutral meson systems is small. If the phase in
the universality violating terms is of ${\cal O}(1)$, a lower
bound on the down squark masses arises, $\mav_{\down}\geq 180\ GeV$.
The effects on the decay $\mu\ra e\gamma$ due to violation of either
 universality or proportionality are more significant and give a
lower bound on the charged slepton masses,
$\mav_{\slepton}\geq 135\ GeV$.
As all sfermion masses are fixed by the gluino mass in this scenario,
we conclude that the most stringent constraint is the one from the
leptonic sector and requires
 $\mg_{3}\geq450\ GeV$ (or equivalently $\mgrav \geq 90\ GeV).$
However, it should be stressed that such estimates are only accurate
up to factors of $\O(1)$.

\section{Constraints from CP Violation}
In the previous section we investigated the effects of violation
of universality and proportionality by string loop effects.
In this section we study then CP violating effects
that arise at the leading order and are implied by eqs.~\FS.
Such effects are
constrained by the upper bounds on the electric dipole moments
of the neutron (EDMN) and of various atoms and molecules.

When both universality and proportionality hold, there are, in general,
two new CP violating phases (in addition to the CKM phase $\delta_{KM}$
and the strong CP phase $\theta_{QCD}$) [\DGH]:
$$
\phi_A\equiv\ \arg\left(A_0\, \mgzero^*\right),\qquad
\phi_B\equiv\ \arg\left(B_0\, \mgzero^*\right),
\eqn\newphases$$
(where $B_0=B_{IJ}/\mu_{IJ}$).
In eq.~\mth\ we defined $\mgrav$ as  a {\it complex}
quantity, its complex conjugate is
$\bar m_{3/2}=\kappa^2 e^{\hat K/2}\hat{\bar W}$.
Using eqs.~\FS\ we conclude that $\phi_A$ vanishes at tree level
while there is  no significant simplification for $\phi_B$ and we
expect [\BIM]
$$\phi_A={\cal O}\left({\alphaX\over4\pi}\right),
\qquad \phi_B={\cal O}\left(1\right)\ .
\eqn\phiAloop$$

The contributions to the EDMs of the neutron and of various atoms
from $\phi_A$ and $\phi_B$ were estimated in ref.~[\FPT].
The appropriate modifications of their estimates to our case read,
in the limit $\phi_B \gg \phi_A$,
\foot{Strictly speaking one should also take the
renormalization of $\phi_B$ into account. However,
in ref.~[\BV] it was shown that $\phi_B$ does not renormalize
and therefore we can use the boundary values at $\MX$.}
$$|d_{\rm N}|\sim 1.4\times10^{-24}\ e\ {\rm cm}\
\left({100\ GeV\over m_{3/2}}\right)^2
\left(\sin\phi_B\right)\eqn\EDMN$$
(where the leading contributions come from the light quark EDMs
and CDMs),
$$|d_{\rm Tl}|\sim 1.6\times10^{-22}\ e\ {\rm cm}\
\left({100\ GeV\over m_{3/2}}\right)^2
\left(\sin\phi_B\right)\eqn\EDMCs$$
(where the leading contribution comes from the EDM of the electron),
and
$$|d_{\rm Hg}|\sim 3\times10^{-26}\ e\ {\rm cm}\
\left({100\ GeV\over m_{3/2}}\right)^2
\left(\sin\phi_B\right)\eqn\EDMHg$$
(where the leading contribution comes from the nonderivative
nucleon-nucleon coupling). The experimental bounds
[\AlSm--\Jaco],
$$\eqalign{|d_{\rm N }|\leq&\ 1.2\times10^{-25}\ e\ {\rm cm},\cr
|d_{\rm Tl}|\leq&\ 6.6\times10^{-24}\ e\ {\rm cm},\cr
|d_{\rm Hg}|\leq&\ 1.3\times10^{-27}\ e\ {\rm cm},\cr}$$
require
$$m_{3/2}\geq480\ GeV\ \sqrt{\sin\phi_B}.\eqn\edmB$$
(We have also checked the bounds from $d_{\rm Cs}$, $d_{\rm Xe}$
and $d_{\rm TlF}$ and found that they are weaker.)
Even for $\sin\phi_B\sim0.1$, we need $m_{\tilde g}\geq800\ GeV$
which is stronger than any of the FCNC bounds \mgq, \mgl.
Finally, we note that if
$\phi_B={\cal O}({\alphaX\over4\pi})$, then \edmB\ is satisfied
for $m_{3/2}\geq30\ GeV$, which
coincides with the direct limit \mbound.

To summarize, of the two new CP violating phases, one
vanishes at string tree level and poses no phenomenological problems.
The other is expected, in general, to be of ${\cal O}(1)$, in which
case it would require gluino mass above $800\ GeV$. Under special
circumstances it could be suppressed,
\foot{For example, when $\tilde\mu=0$ and $\del_S \hat W/\hat{W}$ is
real. Different mechanisms are proposed in refs.~[\Japb, \choi].}
but we see no simple mechanism to guarantee its vanishing.
%%%%%%%%%%%%%%%%%%%%%%%%%%%
%
\chapter{Supersymmetry Breaking induced by the  Moduli}
\section{Non-Universal Soft Terms}
If the dominant source of supersymmetry breaking are moduli $F$-terms
$\vev{F^i}$, the soft scalar masses are generically non-universal
at the string tree level and of $\O(\mgrav)$. The $A$-terms are
not proportional to the Yukawa couplings and of $\O(\mgrav)$.
The gaugino masses are also non-universal
but, more importantly, they are suppressed since $F^S \approx 0$
implies $\mgzero\approx 0$  via eq.~\mgapprox.   Instead we have
$$
\mg_a = {\alphaX\over 4\pi}\, \mg_a^{(1)}  \ ,
\eqn\msuppression$$
where $\mg_a^{(1)} = \O(\mgrav)$. The current lower bound on the
gluino mass [\PDG] implies
$$
150\ GeV < \mg_3 (\Mweak) = 3\, \mg_3 (\MX)
\simeq 3\, {\alphaX\over 4\pi}\, \mgrav\ ,
\eqn\gluinobound$$
or equivalently
$\mgrav \gsim 15\ TeV$.\foot{A similar bound follows from
the charginos. Again, this bound is correct only up to factors of
$\O(1)$ and in specific models a smaller $\mgrav$ might appear [\CCM].}
 Generically, this leads to   squark and
slepton masses of the same order of magnitude
and hence radiative electroweak symmetry breaking
requires a major fine-tuning in order to keep $\Mweak$ at 90 GeV [\BaGi].
However, large scalar masses also suppress the contributions to
FCNC processes.
For a small ratio $\mg_3^2/\msf^2\simeq10^{-4}$ (which follows from
eq.~\gluinobound), the experimental constraints slightly change
compared to \KBDmixMM\ and \megMM\ and now read
$$
\eqalign{
{\rm Re}\VEV{\delta^d_{12}}\leq 1.3 \times10^{-2}
\left({\mav_{\down}\over1\ TeV}\right),&\quad
{\rm Re}(\delta^d_{LR})_{12}\leq 5\times10^{-3}
\left({\mav_{\down}\over1\ TeV}\right);\cr
{\rm Im}\VEV{\delta^d_{12}}\leq 1.1\times10^{-3}
\left({\mav_{\down}\over1\ TeV}\right),&\quad
{\rm Im}(\delta^d_{LR})_{12}\leq 4\times10^{-4}
\left({\mav_{\down}\over1\ TeV}\right),\cr
}\eqn\KBDmixMMm$$
from $\Delta m_K$ and $\epsilon_K$, and
$$
(\delta^\ell_{MM})_{12}\leq 2\times10^{-2}\left({\mav_{\slepton}
\over0.3\ TeV}\right)^2,\quad
(\delta^\ell_{LR})_{12}\leq 1\times10^{-4}\left({\mav_{\slepton}
\over0.3\ TeV}\right),
\eqn\megMMm$$
{}from the bound on BR$(\mu\rightarrow e\gamma)$.
For large scalar masses, the strongest constraint arises in the
down-squark sector from $K-\bar K$ mixing.
(The slepton constraint becomes weaker due to its scaling behaviour.)
 For $\mg^2_a \ll \msf^2$ the $\delta$'s (defined in
eq.~\deltaf) do not renormalize and are given
directly by their boundary values at $\MX$.\foot{
Non-proportional $A$-terms can renormalize the $\delta$'s and
weaken the constraints
but this mechanism is not available for $\mg^2_a \ll \msf^2$ [\CEKLP].}
Off-diagonal scalar mass matrix elements of $\O(\mgrav)$ then imply
$\vev{\delta^d_{12}} \simeq  1$ and hence
 $ \mgrav >\ 75\ TeV$ (or even $\mgrav>\ 650\ TeV$
if ${\rm Im}\vev{\delta^d_{12}}\sim{\rm Re}\vev{\delta^d_{12}}$)
is required in order to satisfy eqs.~\KBDmixMMm. This bound
is much  stronger than the direct bound \gluinobound.

To summarize, for   supersymmetry breaking induced by moduli the
gaugino masses are suppressed and the experimental bound on the gluino
implies rather large squark and slepton masses.
At the same time flavor non-diagonal soft terms are present already
at the string tree level  and despite the large scalar masses they
violate the FCNC bounds.
Thus, one needs at least an approximate universality at leading order.

\section{Universal Soft Terms}
In the moduli dominated scenario universal soft terms appear at
leading order whenever  the couplings $Z_{I\Jb}^{(0)}$ satisfy
$$
Z_{I\Jb}^{(0)} = \hh(\mod,\modb)\  \delta_{I\Jb}\ .
\eqn\Zuni$$
The unit matrix $\delta_{I\Jb}$ in eq.~\Zuni\
is not the only solution which guarantees universal soft terms.
Rather, there could be an arbitrary matrix which only depends on
moduli whose $F$-terms vanish but which is independent on all moduli
whose $F$-terms break supersymmetry. Indeed, $Z_{I\Jb}^{(0)}$ obeys
such a `split'  in string vacua based on $(2,2)$
compactifications where the metric for the ${\bf 27}$ (of $E_6$)
only depends on the $(1,2)$ moduli through an overall
scale factor [\DKL]. (Similarly, the metric for the ${\bf \overline{27}}$
only depends on the $(1,1)$ moduli through an overall scale factor.)

Using eqs.~\RGamma\ and \Zuni\  we find
$$
\Gamma_{i I}^J = \delta_I^J\ \del_i \ln \hh\ ,\qquad
R_{i\jb I\Jb} = Z_{I\Jb}^{(0)}\ \del_i \del_\jb \ln \hh\ .
\eqn\Runi$$
Inserting into \MsIJ\ and \AIJL\ results in
$$
\eqalign{
\mzero\ =&\  |\mgrav|^2 - F^i \Fb^{\jb} \del_i \del_\jb \ln \hh \
\sim \O(\mgrav^2)\ ,\cr
A_{IJL}\ =&\  F^i \left(\del_i Y_{IJL} + Y_{IJL}(\half \del_i \hat{K}
 - 3 \del_i \ln \hh)\right)\ \sim \O(\mgrav Y)}
\eqn\Suni$$
at the leading order (string tree level).
For Yukawa couplings which only depend weakly on the
supersymmetry breaking moduli (\ie\ $\del_i \tilde{Y}_{IJL}\approx0$)
the $A$-terms are strictly  proportional to the Yukawa couplings
($A_0 = e^{\hat{K}/2} F^i (\del_i \hat{K} - 3 \del_i \ln \hh)$).
However, similar to the dilaton case this universality
might be lost at the next order for generic $Z_{I\Jb}^{(1)}$
couplings which do not obey \Zuni\
and we can estimate the physical consequences
implied by such non-universality.
The main difference is that now the gaugino masses
are much smaller than the scalar masses
$\mg^2_a \ll \mzero$ and therefore no renormalization effects enter
into the low energy scalar masses; they are directly determined by
their boundary value $\mzero$. \foot{
In the dilaton-dominated scenario the renormalization of the
squark and slepton masses are driven by the gaugino masses
which is the reason for eqs.~\squarks, \sleptons.}
Similarly, the $\delta$'s do not
renormalize and for both sleptons and squarks we have
$$
(\delta^f_{MM})_{12}\simeq {\alphaX\over4\pi}
{m^{2\, (1)}_{12}\over \mzero}\simeq 3.3\times  10^{-3},
\eqn\deltaqm$$
where we  used  $m^{2\, (1)}_{12}\simeq \mzero$.
(The $(\delta^f_{LR})_{12}$ are  suppressed by an additional
factor of the appropriate fermion mass divided by $\msf$
and thus provide no additional constraint.)
Comparing the theoretical prediction (eq.~\deltaqm) with the
experimental bounds \KBDmixMMm, \megMMm, we see that due to the large
squark and slepton masses implied by the direct limits \gluinobound\
all FCNC constraints are automatically satisfied.

Finally, let us discuss a specific example of the moduli dominated
scenarios which is closely related to no-scale models [\noscale].
For the special case of
$$\hh=e^{\hat K/3}\eqn\LRZIJ$$
in eq.~\Zuni\  (which can also be found in $(2,2)$ vacua),
\Runi\ and  \Suni\ imply\foot{Note that this does not require any
constraint on $\hat K$ itself.}
$$
\eqalign{
\Gamma^N_{iI}=&{1\over3}\delta^N_I\hat K_i\ ,\quad
R_{i\jb   I\bar J}={1\over3}\hat K_{i\jb  }Z_{I\bar J}\ ,\cr
m^2_0 =&0\ , \qquad A_{IJL}=e^{\hat K/2}F^i\partial_i\tilde Y_{IJL}\ .}
\eqn\LRRGamma$$
If, in addition, the moduli dependence
of the Yukawa couplings is weak, $\partial_i\tilde Y_{IJL}\approx0$,
the $A_{IJL}$ terms also vanish at tree level and we have instead
$$A_{IJL}={\cal O}({\alphaX\over4\pi}m_{3/2} ).\eqn\LRAIJL$$
Inserting eqs.~\LRZIJ\ and \LRRGamma\ into eq.~\muBIJ\ gives, in
general, no special cancellations for $B_{IJ}$.
Note, however, that if $H_{IJ}\simeq 0$, then the scale of
$B_{IJ}$ is set by $\tilde\mu_{IJ}$ which is independent of
$m_{3/2}$. In such a case, $B_{IJ}$ could be much smaller
than $m_{3/2}^2$ independently of the SUSY breaking mechanism.
Therefore, in our analysis below, we allow $B_{IJ}$ to
take arbitrary values (as long as they are phenomenologically
acceptable).\foot{$B_{IJ} = 0$ can also be arranged by choosing
$\hat K$ appropriately [\noscale].}

The bound \gluinobound\ still holds but
now the scalar masses also vanish at leading order and one expects
$$
m^2_{I\Jb} \simeq {\alpha_X\over 4 \pi} \mgrav^2 >(850\, GeV)^2 .
\eqn\news$$
Hence,  most parameters in the observable sector are decoupled from
$\mgrav$ at the leading order and only arise from string loop effects
and with the appropriate suppression. However, the contributions
to FCNC processes are generically too large in this scenario.
The bounds are somewhat different from \KBDmixMMm\ and \megMMm\
because in this case $\mg_3^2/\mav_f^2 = 10^{-2}$:
$$
\eqalign{
{\rm Re}\VEV{\delta^d_{12}}\leq 1 \times10^{-2}
\left({\mav_{\down}\over1\ TeV}\right),&\quad
{\rm Re}(\delta^d_{LR})_{12}\leq 6\times10^{-3}
\left({\mav_{\down}\over1\ TeV}\right);\cr
{\rm Im}\VEV{\delta^d_{12}}\leq 8\times10^{-4}
\left({\mav_{\down}\over1\ TeV}\right),&\quad
{\rm Im}(\delta^d_{LR})_{12}\leq 5\times10^{-4}
\left({\mav_{\down}\over1\ TeV}\right),\cr
}\eqn\KBDmixMMn$$
$$
(\delta^\ell_{MM})_{12}\leq 1\times10^{-1}\left({\mav_{\slepton}
\over0.3\ TeV}\right)^2,\quad
(\delta^\ell_{LR})_{12}\leq 1\times10^{-5}\left({\mav_{\slepton}
\over0.3\ TeV}\right).
\eqn\megMMn$$
The strongest constraint again arises in the
down-squark sector from $K-\bar K$ mixing.
For off-diagonal mass matrix elements  of the same order as the
average scalar masses, one has $(\delta^f_{MM})_{12} =\O(1)$ which
implies $m_{\down}>\ 100\ TeV$ (or even $m_{\down}>\ 1000\ TeV$
if ${\rm Im}\vev{\delta^d_{12}}\sim{\rm Re}\vev{\delta^d_{12}}$).

The bounds on $\mgrav$ from electric dipole moments are of $\O(1\
TeV)$ for phases of $\O(1)$. Thus, with $\mgrav\geq\O(10\ TeV)$
these bounds are always satisfied.

\section{Cosmological implications}
The existence of light moduli, $M_{i}\sim M_Z$, with couplings to
observable particles of order $1/m_P$, poses severe cosmological
problems [\CFKRG--\CCQR]. Such moduli are likely to dominate the
matter density of the universe until their decay. When they
decay, at time $\tau_{i}\sim m_P^2/M_{i}^3$, they give a reheat
temperature $T_R\sim\sqrt{m_P/\tau_{i}}\sim10^{-6}\ GeV$,
too low for successful nucleosynthesis ($T_{NS}\sim10^{-3}\ GeV$).

The cosmological implications of the moduli are drastically
different if their masses are much higher than $M_Z$.
A particularly interesting range is $M_{i}\sim {\rm tens\ of}\ TeV$.
If this it the typical mass scale of moduli then [\BCMN,\RaTo]
\item{1.} The universe becomes matter dominated by the heavy moduli
long before they decay if the Hubble constant
during inflation is larger than the moduli masses.
\item{2.} The moduli would decay at time $\tau_{i}\sim
{m_P^2\over M_{i}^3}\sim1\ sec$. They will give a reheat temperature
of $T_R\sim\sqrt{m_P/\tau_{i}}\sim{\rm a\ few}\ MeV$,
just right for nucleosynthesis.
\item{3.} All decay products will thermalize very fast: with typical
number density $n_P\sim T_R^4/M_{i}\sim10^{-16}\ GeV^3$,
hadronic-interaction cross section $\sigma\sim1/f_\pi^2$ and
initial velocity $v\sim1$, the thermalization rate $\sigma n_P v\sim
10^{-14}\ GeV$ is much faster than the expansion rate.
\item{4.} Upon thermalization, the number of photons increases to
$n_\gamma\sim10^{-9}\ GeV^3$, but (as baryon multiplicity in hadron
scattering is ${\cal O}(1-10)$) the number of baryons remains
essentially unchanged, $n_{B+\bar B}\sim n_P\sim10^{-7}n_\gamma$.
(If CP- and B-violating interactions -- either directly in moduli
couplings or indirectly in SUSY interactions -- induce an asymmetry
${n_B-n_{\bar B}\over n_B+n_{\bar B}}\sim10^{-3}$, it would lead
to the required baryon symmetry. However, such a large asymmetry
is unlikely, as a suppression factor $\leq\O({\alpha_s\over\pi})$
is unavoidable.)

Thus, while light moduli pose serious problems to nucleosynthesis,
heavy moduli ($M_{i}\sim100\ TeV$) could actually be {\it responsible}
to nucleosynthesis. \foot{For another solution of the cosmological
moduli problem, that does not require heavy moduli, see ref. [\RaTo].}
{}From eqs.~\msuppression, \LRAIJL\ and \news, we
learn that when (a) SUSY is broken by the moduli, (b) $Z_{I\Jb}^{(0)}=
e^{\hat K/3}\delta_{I\Jb}$,
and (c) the moduli dependence of the Yukawa couplings is weak, then
$\mg_a,\ A_{IJL}\ \ll\ m_{\tilde f}\ \ll \mgrav$ while
the moduli masses are $M_i=\O(\mgrav)$.\foot{ The moduli masses can
only be much lower than $\mgrav$ for special $\hat K$ and $\hat W$
[\noscale].} The direct experimental bound on $\mg_3$ implies then
$$
\mg_3\gsim150\ GeV,\ \ m_{\tilde f}\gsim900\ GeV,\ \ M_i\gsim15\ TeV,
\eqn\hierarchy$$
In this scenario, the moduli masses are necessarily heavy and
consequently the cosmological problems related to light moduli
can be evaded. However, the model faces two problems. First,
in the previous section we found that if universality is violated
at the string one-loop level, then $m_{\tilde f}$ (and consequently
all other scales) should be at least two orders of magnitude
above the bound \hierarchy. In this case, a major fine-tuning
(of order ${M_Z^2\over m_{\tilde f}^2}\sim10^{-6}$) is required
to produce the correct electroweak breaking scale, making
this scenario very unattractive. In order that it
remains viable, there should exist a mechanism that would guarantee
universality to high enough string loop level
that the various scales actually reside not far above the
lower bounds \hierarchy. Second, even if such a mechanism does exist,
the natural scale for $M_Z$ would still be of $\O(m_{\tilde f})$.
We were able to show, however, that with fine-tuning of
$\O({\alphaX\over 4\pi})$ (of either $m_t$ or  $m_{\tilde t}$)
and $\mu = \O({\alphaX\over 4\pi} \mgrav),
B=\O({\alphaX\over4\pi}\mgrav^2)$ we get the correct scale for $M_Z$.
%%%%%%%%%%%%%
\chapter{Conclusion}
In this paper we analyzed the effects of string loop corrections on
rare processes at the weak scale. Since only limited information about
these corrections is currently available, we estimated their
typical order of magnitude and compared them with the stringent
bounds implied by the small FCNC.

We find that in the dilaton scenario the experimental bounds can only
be satisfied by raising the supersymmetry breaking scale,
$$\mg_3\gsim450\ GeV,$$
which is a factor of 3 above the scale
required by the direct experimental limits.
For CP-violating phases of $\O(1)$, constraints from EDMN require an
even larger scale,
$$\mg_3\gsim2.4\ TeV\ \sqrt{\sin\phi_B}.$$
However, both estimates neglect factors of $\O(1)$.
In the moduli scenario the gaugino masses always only
appear as string loop corrections and therefore are
hierarchically smaller than  the scalar masses,
$$\mg_3\gsim150\ GeV,\ \ \msf\gsim15\ TeV.$$
Even with this hierarchy, generic tree level soft terms
violate the bounds from rare processes. For squarks to have their
masses at the lower bound, $\msquark\sim15\ TeV$, the
soft scalar masses that appear at the string tree level
have to be universal. Such universality
does occur with extra conditions on the metric $Z_{I\Jb}$.

We have not explicitly considered the case
where moduli and dilaton $F$-terms are of the
same order of magnitude $F^S \sim F^i$. If eq.~\Zuni\ holds, the soft
parameters  are essentially equivalent to the standard
MSSM parameters at leading order with independent
$\mgzero, \mzero, \mu, A_0, B$. The gaugino masses are not suppressed
and therefore they drive the renormalization of the scalar masses.
Without repeating the entire analysis we may conclude that,
within the accuracy of our estimates, this leads to similar
constraints as were found in the dilaton scenario.
That is, the scale of supersymmetry breaking has to be raised
compared the to scale required by the direct experimental limits.

 In no-scale type scenarios also the scalar masses only
appear at the loop level and
$$\mg_3\gsim150\ GeV,\ \ \msf\gsim900\ GeV,\ \ \mgrav\gsim15\ TeV.$$
This leads to the interesting possibility of a
large hierarchy between the observable sparticle masses
and the moduli masses with interesting cosmological consequences.
However, FCNC constraints push the scale to at least two orders
of magnitude above the lower bounds. For this scenario
to be realistic, universality has to hold well beyond the
string one-loop level.

\ack

We  thank M.~Dine and N.~Seiberg for initiating
this investigation and  T.~Banks, L.~Dixon,
F.~Eberlein, L.~Ib\`a\~nez, A.~K\"onig,
S.~Pokorski and S.~Thomas for useful discussions.
Y.N.~is an incumbent of the Ruth E.~Recu Career Development chair,
and is supported in part by the Israel Commission for Basic Research,
by the United States -- Israel Binational Science Foundation (BSF),
and by the Minerva Foundation.
J.L.~ is supported by a Heisenberg fellowship of the DFG and would
like to thank the Weizmann Institute and Einstein Center
for hospitality and financial support.

\refout

\end